\documentstyle[aps,prb,epsfig]{revtex}
\def\csixty{C$_{60}$}
\def\kfour{K$_4$C$_{60}$}
\def\rbfour{Rb$_4$C$_{60}$}
\def\csfour{Cs$_4$C$_{60}$}
\def\afour{A$_4$C$_{60}$}
\def\athree{A$_3$C$_{60}$}
\def\kthree{K$_3$C$_{60}$}
\def\rbthree{Rb$_3$C$_{60}$}

\def\tuorb{t$_{1u}$}
\def\tuvib{T$_{1u}$}

\def\deg{$^\circ$}

\begin{document}
\draft

\preprint{\today }
%
%

\title{Ordered low-temperature structure in K$_4$C$_{60}$ detected by
infrared spectroscopy}

\author{K. Kamar\'as\cite{kati} and G. Klupp}

\address{Research Institute for Solid State Physics and Optics\\
Hungarian Academy of Sciences \\
P. O. Box 49, Budapest, Hungary, H 1525}

\author{D.B. Tanner and A.F. Hebard}
\address{Department of Physics, University of Florida \\
Gainesville, FL 32611}

\author{N.M. Nemes and J.E. Fischer}
\address{Laboratory on the Research of the Structure of Matter\\
University of Pennsylvania, Philadelphia, PA 19104}

\twocolumn[\hsize\textwidth\columnwidth\hsize\csname@twocolumnfalse\endcsname

\maketitle

\begin{abstract}
Infrared spectra of a K$_4$C$_{60}$ single-phase thin film have
been measured between room temperature and 20 K. At low
temperatures, the two high-frequency T$_{1u}$ modes appear as
triplets, indicating a static D$_{2h}$ crystal-field stabilized
Jahn-Teller distortion of the C$_{60}^{4-}$ anions. The
T$_{1u}$(4) mode changes into the known doublet above 250 K, a
pattern which could have three origins: a dynamic Jahn-Teller
effect, static disorder between ``staggered" anions, or a phase
transition from an orientationally-ordered phase to one where
molecular motion is significant.

\end{abstract}
]

The electronic structure of fulleride salts has been a topic of
intense interest since their discovery. Three recent reviews
summarize the situation: Reed and Bolskar\cite{reed}\ focus on
isolated ions; Forr\'o and Mih\'aly \cite{lacilaci}\ treat
fulleride solids; and Gunnarson \cite{gunnar} concentrates on the
theory of superconductivity and on band structure calculations,
including the important concepts of the Jahn-Teller (JT)
effect.\cite{chanc} Fulleride anions with charge other than 0 or 6
are susceptible to Jahn-Teller coupling between the electrons
occupying partially-filled \tuorb\ orbitals and H$_g$ phonons.
Because of the high degeneracy of the phonons involved, the
coupled electronic Hamiltonian acquires SO(3) symmetry, higher
than the original icosahedral configuration.\cite{pooler} Thus the
electronic structure of the outer shell can be discussed in
complete analogy with {\it p}-electrons on an atom.\cite{lacilaci}
In any uniaxial crystal field, one expects a twofold orbital
splitting to e$_{2u}$ and a$_u$, whereas in a biaxial system the
degeneracy is completely lifted, resulting in three distinct
electronic levels. (It follows from the SO(3) symmetry that the
principal axes of the distortions can be taken as the axes of the
crystallographic unit cell, regardless of the position of the
molecule within the crystal.)

In solids the orbital splitting is small compared to the
bandwidth; thus, electron correlation has to be taken into
account. The most comprehensive theoretical treatment of this
interplay was given by Tosatti and co-workers, who introduced the
concept of the ``Mott-Jahn-Teller insulator" (M-JT).\cite{fab,cap}
The model of {\it p}-like electrons on a spherical or ellipsoidal
surface was indeed succesful in the interpretation of
NMR\cite{zimmer,kerkoud}, photoelectron
spectroscopy,\cite{knupf,finkb} resonance Raman\cite{ruani} and
magnetic\cite{luk} data. The situation becomes different, however,
once we try to measure properties that reflect the internal
degrees of freedom within the C$_{60}$ ball itself. Typical
examples are magnetic resonance measurements and vibrational
spectroscopy. When atomic motion has to be taken into account, the
symmetry can no longer be regarded as spherical. Instead, the
actual shape of the molecule is determined by the combined
internal (JT) and external (crystal field) distorting forces. If
these result in a potential surface with several shallow wells,
then the balls will assume different shapes and positions with
respect to the lattice. The result will be either static disorder
or random fluctuations between the minima, called the dynamic
Jahn-Teller effect.

A special frustration results if the principal axis of the crystal
system does not coincide with a similar molecular axis. The case
of A$_4$C$_{60}$ salts \cite{flem} is such an example. These show
peculiarities in both their crystal structure and vibrational
spectra. At room temperature, K$_4$C$_{60}$ and Rb$_4$C$_{60}$
form a bct crystal,\cite{knutsch} with merohedral disorder similar
to that in \athree:\cite{steph} in the bct cell of \kfour\ there
is only a $C_2$ molecular axis in the $c$ direction, and the
overall tetragonal structure is realized by the anions occupying
two standard orientations with equal probability. In
Cs$_4$C$_{60}$ an orthorhombic distortion was found at room
temperature,\cite{dahlke} resulting in a lattice compatible with
D$_{2h}$, one of the subgroups of the icosahedral group yielding
complete ordering of the \csixty\ balls.

Infrared measurements on \kfour\ and \rbfour\ at room temperature
by Iwasa and Kaneyasu \cite{iw1} found the T$_{1u}$(4) infrared
active mode to  be split into a doublet, a phenomenon not seen in
any other alkali salt so far (excluding polymeric forms, which
show a variety of splitting patterns.) They invoked a static JT
distortion to explain these results. X-ray
diffraction\cite{knutsch} puts an upper limit of 0.04 \AA\ for the
difference between the polar and equatorial radii of \csixty\ (the
polar direction supposed to lie in the crystallographic
$c$-direction). To our knowledge, neither structural nor
vibrational properties of the \afour\ salts have been reported at
low temperatures.

Details concerning the temperature dependence of the molecular
dynamics in \afour\ salts were provided by nuclear magnetic
resonance studies.\cite{zimmer,kerkoud,zimmerk,rachdi} Goze et
al.\cite{rachdi} observed four $^{13}$C NMR lines in \csfour\
collapse into a single line above 350 K, the latter spectrum being
similar to \kfour\ at room temperature.\cite{zimmerk} Lacking
structural data, they could not draw unambigous conclusions, but
they invoked the possibility of a phase transition between an
ordered and a rotationally disordered phase and even suggested
that a similar transition may take place in other tetravalent
alkali salts at lower temperatures. Comparing the NMR results with
high-quality room-temperature x-ray diffraction data obtained in
the meantime on \kfour\ and \rbfour\ (which are
isostructural)\cite{knutsch} and \csfour,\cite{dahlke} it seems
highly possible that a phase transition occurs in all three salts,
above room temperature in \csfour\ and below in the other two. Due
to its sensitivity to local symmetry, infrared vibrational
spectroscopy should be a good indicator for such a transition.
Therefore, we undertook infrared spectroscopic measurements
between 20 and 300 K in \kfour.

Highly polycrystalline films of \csixty\ were grown on Si IR
windows using a ``hot-wall" technique.\cite{fischer} We estimate a
thickness of 300 nm from the spacing of interference fringes in
the infrared, assuming an index of refraction $n=2$. A glass and
metal doping cell was constructed incorporating a potassium source
and ZrO getter (SAES Getters/ USA Inc.), replicating earlier
preparations from Bell Labs.\cite{hebard} The cell was thoroughly
outgassed at 300$^{\circ}$C and sealed at a base pressure of
$10^{-7}$~torr. During doping the cell was kept at 200$^{\circ}$C
and the resistance was monitored. The minimum resistance
corresponding to \kthree\ was observed and the time required to
reach \kfour\ composition estimated. Then the sample was
transferred in an Ar filled glovebox into a copper holder and
epoxied in place.

Spectra were taken in the midinfrared by a Bruker IFS28
Fourier-transform infrared spectrometer with an MCT detector and
in both the far and midinfrared by a Bruker IFS113v instrument
using a Si bolometer and a DTGS detector, respectively.
Temperature dependence was measured in a He flow cryostat while
warming up, to avoid possible hysteresis effects. Although the
measurements were taken in reflection mode with the silicon window
facing the incident light path, our results are essentially
transmittance data. This is because the experimental arrangement
resulted in the light passing through the silicon window and the
sample film twice, once on entering the sample and a second time
after being reflected from the copper back plate. Because we were
mainly interested in the change in spectral line positions, we did
not undertake any correction for reflection from the silicon
window, and assumed 100\% reflection from the copper.

\begin{figure}
\epsfig{file=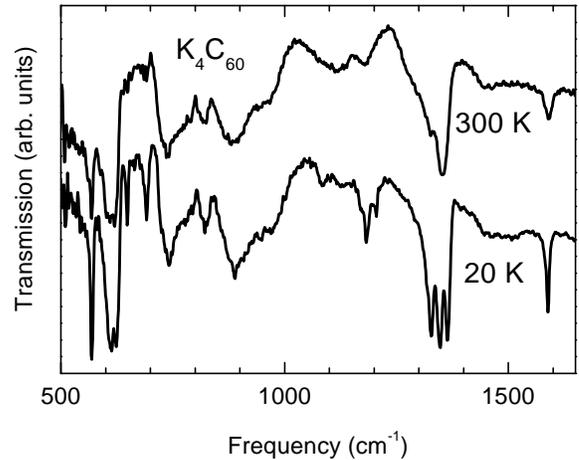, clip=true}
\caption{Infrared spectra of a \kfour\
layer on silicon in the entire midinfrared range at 20 and 300 K.}
\label{one}
\end{figure}

Figure 1 shows the infrared spectrum of the film at 300 K and at
20 K. The 300 K spectrum is identical to that measured by Iwasa
and Kaneyasu,\cite{iw1} the most pronounced feature being the
asymmetric doublet at 1323 and 1353 cm$^{-1}$. The \tuvib(3) mode
at 1177 cm$^{-1}$ is broader than that in other fulleride salts,
but shows no explicit splitting. Neither of these peaks coincides
with infrared vibrations of other fulleride ions; moreover,
characteristic peaks of other fulleride ions are absent, giving
evidence for the phase purity of the film. The low-frequency
T$_{1u}$ mode region is obscured by  multiphonon Si absorption
occurring between 570 and 700 cm$^{-1}$, therefore we concentrate
on the high-frequency region, where neutral isolated C$_{60}$ has
two (triply degenerate) normal modes  at 1183 and 1429 cm$^{-1}$.
In Figure 2, this part of the spectrum is depicted at 20, 200, and
300~K. An obvious change occurs between 300 K and 200~K; at 20 K
the structure becomes sharper, but otherwise the spectrum is
identical to that at 200 K. At our heating rate of 2 K/min, we
observed this change between 250 and 260 K. The two high-frequency
\tuvib\ modes both evolve into a triplet below this temperature (a
set at 1168, 1184 and 1206, and another at 1327, 1347 and 1364
cm$^{-1}$).

The threefold splitting in the \tuvib\ peaks is a clear signature
of biaxial symmetry, a complete lifting of the threefold
degeneracy in the vibrational levels. A straightforward
explanation is the existence of an ordered crystal structure at
low temperature, similar to that of \csfour.\cite{dahlke} There,
the site symmetry of the balls is D$_{2h}$, a direct subgroup of
I$_h$. This subgroup is also special in the sense that it can
describe one possible shape of a JT distorted icosahedral
molecule, resulting from linear coupling between \tuorb\ electrons
and H$_g$ phonons.\cite{chanc} The principal axis of the point
group coincides with that of the crystal space group, thus all the
conditions for a crystal-field stabilized static JT distortion are
met. We conclude that the likely explanation for the
low-temperature infrared spectrum of \kfour\ is such a structure.

\begin{figure}
\epsfig{file=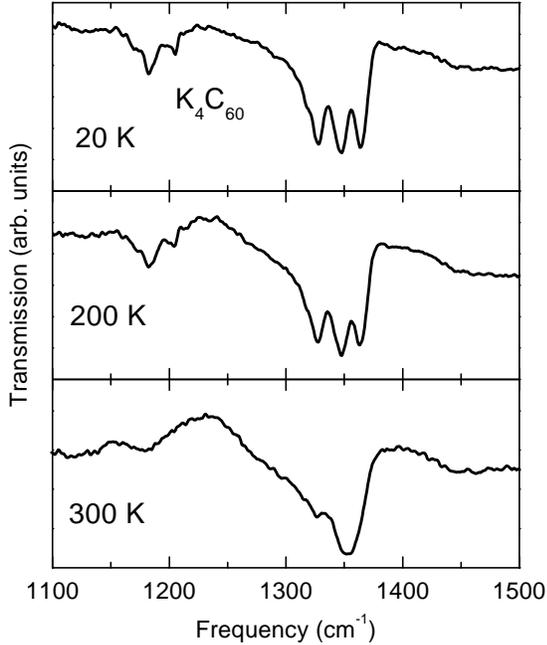, clip=true}
\caption{Infrared spectra of \kfour\ in
the region of the high-frequency \csixty\ molecular vibrations, at
three different temperatures.}
\label{two}
\end{figure}

It is not straightforward to interpret the spectrum at 300 K,
where the structure is known. If we assume a static merohedral
disorder with appropriate JT distortion, then molecules with
different standard orientations should give the vibrational
spectrum of the local symmetry, D$_{2h}$, and thus should not
differ from the low temperature spectra. Because the spectra do
differ, we suggest that either the molecules are trapped in more
complicated orientations (the ``staggered static distortion"
introduced by Fabrizio and Tosatti\cite{fab}), or that dynamical
processes are involved. The two allowed distorted point groups for
\tuorb-H$_g$ coupling besides D$_{2h}$ are D$_{3d}$ and D$_{5d}$.
The principal axes of these two distorted symmetries go
respectively through the center of the hexagonal and pentagonal
faces of the C$_{60}$ molecule. The staggered static distortion
would have neighboring anions with principal axes different from
$c$, but symmetric around it; no experimental corroboration of
this idea has been reported. It is interesting to compare the
measured upper limit for the static distortion, 0.04
\AA,\cite{knutsch} with the only hitherto known \csixty\ distorted
structure, that of C$_{60}^{2-}$ in (PPN)$_2$C$_{60}$.\cite{paul}
(The JT coupling in this ion is supposed to be very similar to the
tetravalent anion, because they are related by electron-hole
symmetry.) There, a distortion of 0.033 \AA ngstroms was enough to
produce a split EPR signal in accordance with symmetry lowering of
the anion,\cite{boyd} but the point group was $C_i$, lower than
that predicted by the JT effect alone (due to the complicated
organic counterions), and involved a considerable D$_{5d}$
character, pointing to a D$_{5d}$ distortion stabilized by crystal
field effects and locked into a configuration with even lower
symmetry.

In the case of \kfour, however, the $C_3$ axes seem more special.
We consider first their role in \athree, which adopts a remarkably
similar molecular arrangement in the $ab$ plane\cite{steph} with
two standard orientations, connected formally by a 90\deg\
rotation around the $c$ axis. In \rbthree\ and \kthree the
rearrangement between the two standard orientations can also occur
through a 44.5\deg\  jump around a $C_3$\ axis, lying in the
crystallographic $\langle 111\rangle$ direction, and it has been
shown by neutron scattering experiments that this motion involves
a considerably lower energy barrier than the jump around
$C_2$.\cite{reznik} Subsequent NMR studies\cite{krizal,krizab}
indicate that a fast uniaxial rotation takes place around a fixed
axis, accompanied by a slow motion which was interpreted as the
flipping of the molecule while keeping the same rotational axis.
The existence of a fast and slow reorientational motion has also
been reported in \afour.\cite{zimmer,zimmerk} If one invokes the
special role of the $C_3$ axes, then it is possible that the easy
rotation around this axis lowers the energy of the D$_{3d}$
distortion over both D$_{5d}$ and D$_{2h}$.

At this point, we turn to the description by Chancey and
O'Brien\cite{chanc} of the dynamical Jahn-Teller effect in
\csixty. They introduce the concept of ``pseudorotation'', a
succession of symmetry-allowed distortions progressing on the
surface of the ball, causing a continuous change in the shape of
the molecule without effectively rotating it. Comparing this model
to that of the motions deduced from NMR spectroscopy, we can
associate the axial rearrangement with such a pseudorotation
(obviously, a genuine rotation should involve changes in the
positions of atoms, because it takes the molecule from one
standard orientation to another). This motion should go through
distorted states which mainly involve axial components (D$_{3d}$
and/or D$_{5d}$) and can be considered as the dynamic equivalent
of the staggered state proposed by Fabrizio and Tosatti.\cite{fab}
Based on vibrational spectra, it is not possible to distinguish
between the dynamic process and the staggered state, so long as
biaxial (D$_{2h}$) distortions are not involved; both the spatial
and time average of uniaxially distorted states would give a
twofold line splitting. The transition between the low- and high-T
states, however, would be different in the two cases. A staggered
tetragonal phase can only evolve from an ordered structure through
a structural phase transition. Such a transition could, of course,
also happen between an ordered and a dynamically disordered phase,
as in pristine \csixty, but it is also possible that the
disordered state develops from the ordered one by the rotational
frequencies increasing above a critical value. The first case
should lead to a discontinuity at the transition temperature
observable by any spectroscopic method, regardless of time scale;
the second scenario would give different transition temperatures
depending on the time scale of the experiment relative to the
autocorrelation time of the motions.\cite{virginia} From the
parameters given by Zimmer et al.\cite{zimmerk}, the
autocorrelation time $\tau$ is 33 ns at 200 K and 1 ns at 300 K,
which can be regarded as static on the time scale of molecular
vibrations. The same calculation for the fast uniaxial rotation
gives 71 ns at 200 K and 130 ps at 300 K, still too slow to cause
collapsing of infrared lines. (In \csixty, $\tau$ = 64 ns at 200
K, below the structural phase transition,\cite{tycko} where
infrared lines are observably split,\cite{homes} and 12 ps at 300
K, in the rotator phase, where there is only a single infrared
peak per \tuvib\ mode.) We conclude therefore that the temperature
evolution of the infrared spectra cannot be explained by simple
activated behavior of some molecular motion; rather, an abrupt
phase transition has to occur to a state where the rotations are
faster. Suggestions for a phase transition have been put forward
by Goldoni et al.\cite{goldoni} based on an anomaly in the C-$1s$
photoemission linewidth below 150 K, which they correlate with the
peak in the NMR spin-spin relaxation time\cite{zimmerk}, and
attributed to the stopping of the slow reorientational motion. We
find this suggestion difficult to reconcile with the results
presented here, in part because our transition temperature is
higher than 150 K, and in part because the splitting of IR lines
should be related to the fast motion rather than the slow
reorientation.

More structural and spectroscopic data are clearly needed on
fulleride salts to fully understand their intriguing behavior. In
addition, vibrational spectra of isolated ions could tell whether
the splitting at either high or low temperature is caused by an
inherent molecular JT distortion or by a crystal field effect
brought about by the surrounding cations. Indirect proof of a
molecular distortion comes again from studies\cite{boyd} on
C$_{60}^{2-}$, where the same EPR structure was found in the PPN
crystal and in frozen solution, indicating that the effect of
crystal structure was not dominant. Therefore, we think it
possible to have a D$_{2h}$ configuration in the low-temperature
state and a combination of the two axially distorted forms,
D$_{5d}$ and D$_{3d}$ connected by a pseudorotation, in the
high-temperature state.

In conclusion, we observed a change in the T$_{1u}$(3)) and
T$_{1u}${(4) infrared-active vibrations of C$_{60}$ in
K$_4$C$_{60}$ from doublet to triplet in the 200-300~K range. We
propose a structural bct-to-orthorhombic transition to take place
in this region.

We thank G. Oszl\'anyi and G. Kriza for enlightening discussions.
This work has been supported by OTKA grants T 34198 and T 29931
and by an NSF-MTA-OTKA International Grant (No. 021, N31622 and
NSF-INT-9902050). Work at Penn supported by NSF DMR97-30298.




\end{document}